\documentclass[preprint,preprintnumbers,showpacs,showkeys,amssymb,amsmath,
  prb,aps]{revtex4}

\usepackage{graphicx}
\usepackage{amssymb}

\begin{document}

\title{Unified Framework for Dislocation-Based Defect Energetics}

\preprint{LA-UR-04-XXXX, MIAM-04-01}

\author{J. M. Rickman}
\affiliation{Department of Materials Science and Engineering,
Lehigh University,
Bethlehem, PA 18015, USA}

\author{Jorge Vi\~{n}als}
\affiliation{McGill Institute for Advanced Materials and Department of Physics,
3600 University St,
Montreal, QC H3A 2T8, Canada}

\author{R. LeSar}
\affiliation{Theoretical Division, Los Alamos National Laboratory,
Los Alamos, NM 87545, USA}
\date{\today}

\begin{abstract}
We present a unified framework for the calculation of defect
energies for those defects that can be represented as a superposition
of isolated dislocations, and obtain both self and interaction
energies of combinations of grain boundaries and cracks. We recover in
special limits several well known quantities such as the energy of 
a low-angle tilt boundary, as well as other lesser known results, 
including boundary/boundary and crack/boundary interaction energies. 
This approach, in combination with simple dimensional analysis,
permits the  rapid calculation of defect energetics in the elastic
limit.
%
\end{abstract}

\pacs{61.72.Bb}
\keywords{Dislocation Density Tensor; Multipole Moments}

\maketitle

\section{Introduction}
The presence of extended defects in crystalline solids, such as grain
boundaries and cracks, can dramatically affect their mechanical
response. For example, in large-grained polycrystalline materials, grain boundaries 
may sometimes impede dislocation motion with a concomitant increase in
yield strength, \cite{cottrell67,bata04} while sharp cracks are stress
concentrators that may initiate material fracture
\cite{bowman04,sridhar95}. From a micromechanical point of view, the
elastic fields associated with grain boundaries or cracks lead 
to defect interactions that influence fracture behavior in 
highly-flawed systems and solute segregation\cite{li,wang04},
thereby affecting strengthening or embrittling mechanisms. 

Calculations of extended defect interaction energies most often begin
with the solution of the appropriate elastic boundary value problem, followed 
by the construction of the corresponding energy density in terms of
the stress (or strain) fields and elastic constants and the
integration of this density over some volume of space. An alternative
approach to the defect interaction problem follows from a 
micromechanical model of the defect and is based in the elastic Green 
function of the medium. Such an approach was pioneered by Mura and
others \cite{mura87,teodosiu}, and will be applied here to defects
that may be regarded, in some limit, as 
composed of elemental straight dislocations.  As will be seen below, this formalism 
permits the straightforward calculation of interaction energies for 
different extended defects and, moreover, facilitates an intuitive, 
multipole-based analysis that reveals their asymptotic dependence on
defect separation.

We address in this paper two prototypical systems: grain boundaries
and cracks, as these defects can often be modeled in terms of spatial
distributions of 
line defects, the former typically by regular array(s) of straight 
dislocations each having a constant Burgers vector, and the latter by 
a continuous, localized distribution of dislocations with a Burgers 
vector density depends upon loading conditions.  Thus, the 
eigenstrains corresponding to each defect are conveniently expressed in terms of a 
dislocation density tensor that embodies the relevant defect length 
scales and separations.  For the purposes of illustration we consider 
the energetics of a low-angle tilt boundary, the interaction between 
tilt boundaries, the energetics of an array of cracks, and the 
interaction between an isolated crack and a grain boundary.  The 
application of this approach to other, related examples is then 
discussed.

\section{Formalism and Selected Illustrations}
As indicated above, our calculations are facilitated by describing a
defect (e.g., grain boundary, crack) in terms of a dislocation density 
tensor, $\bar{\rho}$ \cite{mura87}. This solenoidal, second-rank
tensor carries both local Burgers vector and line direction
information, with a divergenceless condition that is a 
consequence of the topological constraint of line continuity.  For 
example, the components of the density tensors corresponding to straight 
edge and screw dislocations at $r_{1}=r_{2}=0$ with Burgers vectors
$\vec{b}$, and line directions along $r_{3}$ are given, respectively, by
\begin{eqnarray}
\label{eq:densities}
\rho_{ij}(\vec{r}) & = & b \; \delta_{i3} \; \delta_{jJ} \; \delta(r_{1}) \; 
\delta(r_{2}), \;\; ({\rm edge},\; J=1,2) \nonumber\\
\rho_{ij}(\vec{r}) & = & b \; \delta_{i3} \; \delta_{j3} \; \delta(r_{1}) \; 
\delta(r_{2}), \;\; ({\rm screw}).
\end{eqnarray}
Note that, in the first equation, it is assumed that the Burgers 
vector is aligned either along $r_{1}$ or $r_{2}$, and so an 
arbitrary alignment in the plane may be constructed from appropriate linear 
combinations. $\delta(r)$ is the Dirac delta function.

The elastic energy of the system $E\left[ \bar{\rho} \right]$ can be
written as a functional of $\bar{\rho}$, as shown by Kosevich and
others \cite{kosevich} and, more recently, has been employed by Nelson and 
Toner \cite{nelson81} to investigate the effect of unbound dislocation
motion on shear 
response in solids, by Rickman and Vi\~{n}als \cite{rickman97} to
model the collective motion of dislocation ensembles, and by Rickman
and LeSar \cite{rickman01} to quantify the temperature dependent interaction of
fluctuating dislocation lines. For our purposes, it is convenient to
express the energy functional for an elastically isotropic medium with
shear modulus $\mu$ and Poisson ratio $\nu$ as the Fourier integral
\begin{equation}
\label{eq:energy1}
E[\bar{\rho}] = \frac{\mu}{2 (2\pi)^{3}} \; \int d^{3}q \; 
\frac{1}{q^{2}} \;  K_{ijkl}(\vec{q}) \; 
\tilde{\rho}_{ij}(\vec{q}) \; \tilde{\rho}_{kl}(-\vec{q}) ,
\end{equation}
where the integration is over reciprocal space (tilde denoting a 
Fourier transform), the kernel (without defect core energy
contributions) is given by
\begin{equation}
\label{eq:kernel1}
K_{ijkl} = \left[Q_{ik} Q_{jl} \, + C_{il}
C_{kj} \, + \, \frac{2 \nu}{1 - \nu} C_{ij} C_{kl} \right] ,
\end{equation}
and $\bar{Q}$ and $\bar{C}$ are longitudinal and transverse projection
operators, defined by
\begin{eqnarray}
\label{eq:qij}
Q_{ij} & = & \delta_{ij} - \frac{q_i q_j}{q^2} \nonumber \\
C_{ij} & = & \epsilon_{ijl} \frac{q_l}{q} \, ,
\end{eqnarray}
respectively \cite{nelson81}, where $\delta_{ij} $ is the Kronecker
delta, and $\epsilon_{ijk}$ is the Levi-Civita tensor.
%
%
We note that it is possible to include defect core energies in a 
somewhat {\it ad hoc} fashion by augmenting this kernel by a term
quadratic in the dislocation density \cite{nelson81}  A somewhat more
realistic description of the core could be given by an energy term
that depends on atomic 
coordinates, although we will not consider such core models as
they fall outside of the mesoscopic description outlined here. As illustrated 
below, the distance dependence of the defect energy can be determined
in a straightforward manner from Eq. (\ref{eq:energy1}) by employing a
superposition of prototypical densities.

\subsection{Edge Dislocation Arrays - Low-Angle Tilt Grain Boundaries}

\begin{figure}
\includegraphics[scale=0.4]{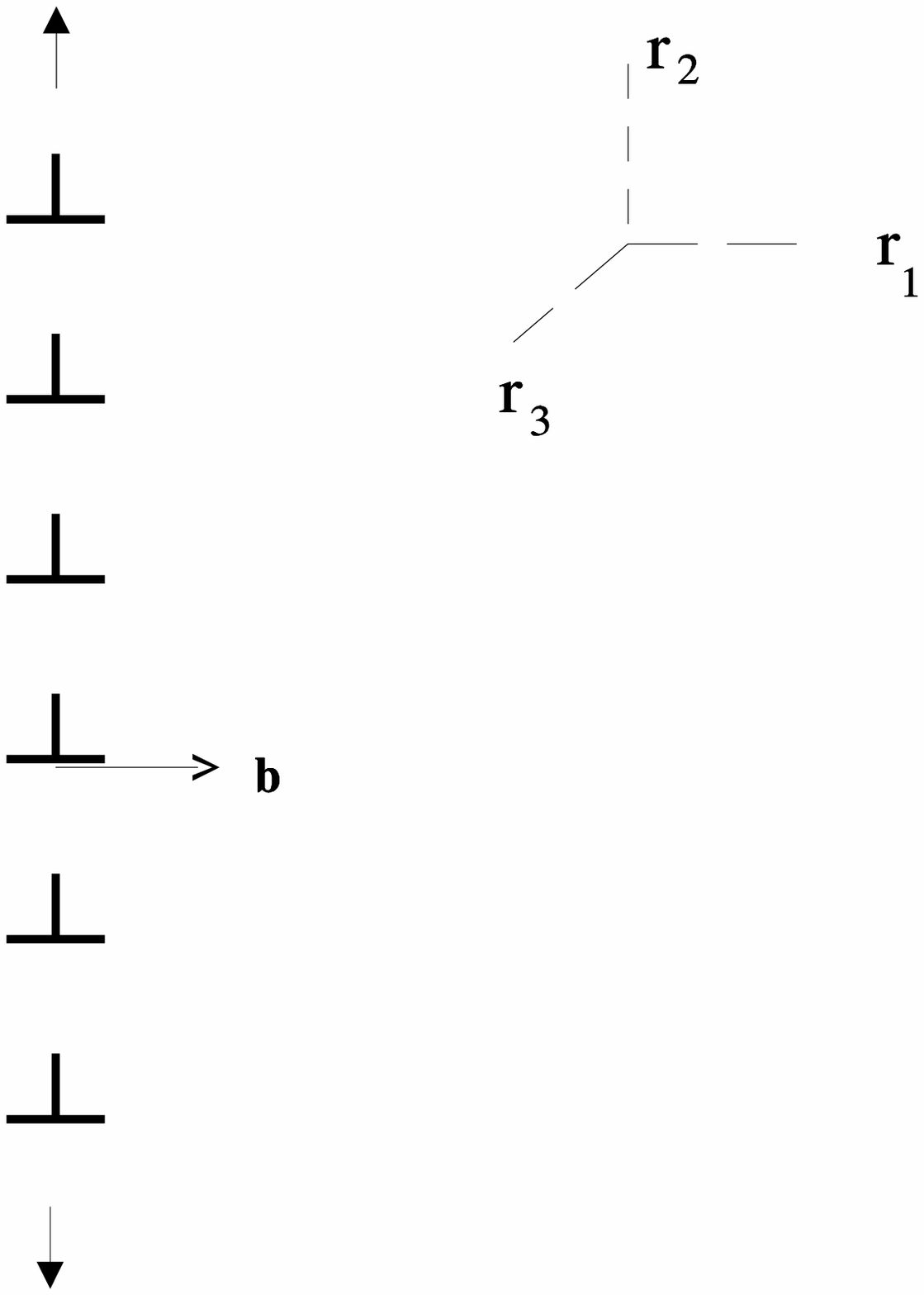}
\includegraphics[scale=0.4]{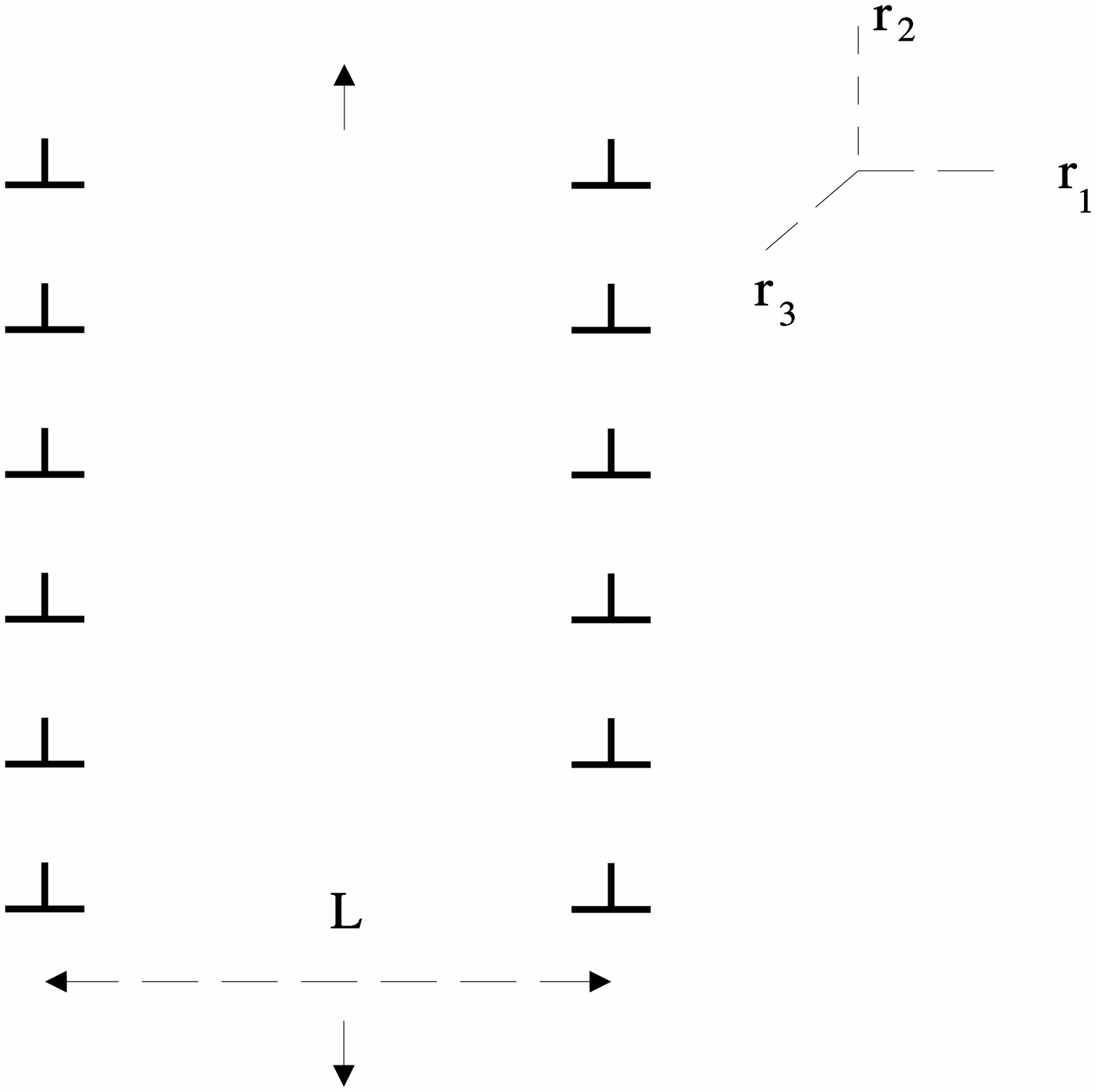}
\caption{ a.) A low-angle, tilt grain boundary modeled as an array of
  edge dislocations.  Each dislocation has a Burgers vector $\vec{b}$
  and is separated from its nearest neighbor by $\ell \approx
  b/\Theta$ for small misorientation angle $\Theta$.  b.)  Two such
  grain boundaries separated by a distance $L$.}
\label{fig:fig1}
\end{figure}

Consider first a linear array of $N$ edge dislocations, each separated 
from its nearest neighbor by $\ell$ and having a Burgers vector aligned 
along either the $r_{1}$ or $r_{2}$ axes ($J$ = 1 or 2, 
respectively), as shown in Fig. $\ref{fig:fig1}$a.  
As is well known, this system, with $J=1$ and in the limit $N
\rightarrow \infty$, is a model of a low-angle tilt grain 
boundary where, in the limit of small grain misorientation angle 
$\Theta$, $\ell \approx b/\Theta$ \cite{hirth}. From Eq. (\ref{eq:energy1}) the 
dislocation density is
\begin{equation}
\label{eq:denstilt}
\rho_{ij}(\vec{r}) = b \; \delta_{i3} \; \delta_{jJ} \; \delta(r_{1})
\; \sum_{n} \delta(r_{2}-n\ell),
\end{equation}
where the summation is over all dislocations in the array.  For 
convenience, we will work with the corresponding Fourier transform
\begin{equation}
\label{eq:denstiltft}
\tilde{\rho}_{ij}(\vec{q})  =  2 \pi b \; \delta(q_{3})
\sum_{n} \exp{( i q_{2} n \ell )}
\; \delta_{i3} \; \delta_{jJ} .
\end{equation}

Before calculating the corresponding boundary energy, we note that
singularities inherent in this continuum model that lacks a short
distance cut-off lead to a divergent self-energy. The origin of the
divergence is the fact that the energy density of a 
single edge dislocation diverges like $1/r$, where $r$ is the radial 
distance to the dislocation line. In reality, each dislocation has an 
atomic core region that is not amenable to a continuum description. We
introduce a small cutoff parameter $a$ representing 
the core radius by replacing the real-space delta function by the
broadened delta representation \cite{alerhand} 
\begin{equation}
\Delta(r_{2}-nl) = \frac{a}{\pi} \; \frac{1}{a^{2}+(r_{2}-n\ell)^{2}}.
\end{equation}
The corresponding Fourier transform can be obtaining by using the 
shifting property of transforms to obtain
\begin{equation}
\label{deltaft}
\tilde{\Delta}(q_{2}) = \exp{(i q_{2} n\ell )} \; \exp{(-a |q_{2}|)}.
\end{equation}
%
%

Therefore the elastic energy of the array of edge dislocations is
\begin{equation}
\label{eq:energy}
E_{J} = \frac{\mu}{2 (2\pi)^{3}} \; \int d^{3}q \; \frac{1}{q^{2}} \;
K_{3J3J}(\vec{q}) \; 
\tilde{\rho}_{3J}(\vec{q}) \; \tilde{\rho}_{3J}(-\vec{q}) ,
\end{equation}
where
\begin{equation}
\label{eq:kernel}
K_{3J3J}(\vec{q}) = Q_{33}Q_{JJ} + \frac{1+\nu}{1-\nu} C_{3J} C_{3J}
\end{equation}
and $J$ = 1 or 2 (no summation) denotes the orientation of the 
Burgers vectors.  More specifically, noting that $q^{2} = q_{1}^{2}+q_{2}^{2}$,
\begin{equation}
\label{kvalues}
K_{3J3J}(\vec{q}) = \left(\frac{2}{1-\nu}\right)  
\frac{q^{2}-q_{J}^{2}}{q_{1}^{2}+q_{2}^{2}},
\end{equation}
and the energy per unit length in the $r_{3}$ direction is given by
\begin{equation}
\label{eq:enstrfac}
e_{J} = \frac{\mu b^{2}}{(2\pi)^{2}\; (1-\nu)} \int \int d^{2}q \;
\frac{q^{2} - q_{J}^{2}}
{\left(q_{1}^{2}+q_{2}^{2}\right)^{2}}\; S(q_{2})
\exp{(-2a|q_{2}|)},
\end{equation}
where a structure factor $S(q_{2})$ has been defined as $S(q_{2}) =
|\sum_{n} \exp{( i q_{2} n \ell )}|^{2} =  
N (2\pi/\ell) \; \sum_{m} \delta(q_{2} - 2\pi m/\ell)$
\cite{kleinert89}. Explicitly we have 
\begin{eqnarray}
\label{eq:ener}
e_{1} & = & \frac{N \mu b^{2}}{(2\pi)\; (1-\nu) \ell} 
\sum_{m=-\infty}^{\infty} \; \exp{(-4\pi a|m|/\ell)} \;
 \int dq_{1} \; \frac{(2\pi m/\ell)^{2}}
{\left(q_{1}^{2}+(2 \pi m/\ell)^{2}\right)^{2}}\;
 , \nonumber\\
e_{2} & = & \frac{N \mu b^{2}}{(2\pi)\; (1-\nu) \ell} 
\sum_{m=-\infty}^{m=\infty} \; \exp{(-4\pi a|m|/\ell)} \;
 \int dq_{1}\; \frac{q_{1}^{2}}
{\left(q_{1}^{2}+(2 \pi m/\ell)^{2}\right)^{2}}\; . \
\end{eqnarray}

Upon evaluating the first integral in Eq. (\ref{eq:ener}),  one finds
\begin{equation}
\label{eq:e1eval}
e_{1} = \frac{ N \mu b^{2}}{2 (2\pi)\; (1-\nu)} 
\sum_{m=1}^{\infty} \; \frac{\exp{(-4\pi a|m|/\ell)}}{|m|},
\end{equation}
and therefore
\begin{equation}
\label{eq:e1cont}
e_{1} = - \frac{ N \mu b^{2}}{(4\pi)\; (1-\nu)} \; \ln{\left[1-\exp{(-4\pi 
a/\ell)}\right]} \approx \frac{ N \mu b^{2}}{(4\pi)\; (1-\nu)} 
\ln{\left[\frac{\ell}{4\pi a}\right]},
\end{equation}
where the latter approximation holds when $a/\ell$ is small.  If this 
array represents a planar defect (i.e., grain boundary) then the 
corresponding energy per unit boundary area is
\begin{equation}
\label{eq:gbener}
\gamma = \frac{e_{1} }{N \ell}  \approx \frac{\mu b^{2}}{(4\pi \ell)\; (1-\nu)} 
\ln{\left[\frac{\ell}{4\pi a}\right]}.
\end{equation}
This result has the same functional form as the classic Read-Shockley 
energy for a low-angle tilt boundary, though the constant in the argument 
of the logarithm depends on the details of the short-range (core) 
cutoff employed here \cite{hirth}.  In this regard, we note that a 
similar approach to this problem for two-dimensional systems has been 
outlined elsewhere \cite{chaikin}.

Finally we examine $e_{2}$ for the extended defect for which $\vec{b} 
\parallel \hat{r}_{2}$.  From Eq. (\ref{eq:ener}) it is clear that  
$e_{2} \rightarrow \infty$ owing to the divergence associated with the 
$m$ = 0 mode.  This result is expected since some components of the 
long-ranged stress field (namely $\sigma_{22}$) tend to a constant 
as $x \rightarrow \infty$, and thus the volume integral over the 
energy density diverges in an infinite system.  Given this behavior, 
it is evident that a low-angle boundary with this geometry cannot have
dislocations with Burgers vector components in the boundary plane.

\subsection{Dislocation Array Interactions}

Consider next two low-angle, symmetric, tilt boundaries separated by a 
distance $L$, as shown in Fig. $\ref{fig:fig1}$b. The interaction
energy for these boundaries follows from the corresponding dislocation
density tensor 
\begin{equation}
\rho_{ij}(\vec{r}) = b \delta_{i3} \; \delta_{j1} \left[ \delta(r_{1}) 
\; \sum_{n} \delta(r_{2}-n\ell) \; + \delta(r_{1}-L) \; \sum_{m} 
\delta(r_{2}-m\ell) \right]
\end{equation}
that can be transformed to yield
\begin{equation}
\label{eq:densintft}
\tilde{\rho}_{ij}(\vec{q}) = (2 \pi b) \delta_{i3} \; \delta_{j1} \; \delta(q_{3})
\left[\sum_{n} \exp{( i q_{2} n \ell )} \; + \; \exp{\left(i q_{1} L \right)} \; 
\sum_{m} \exp{(i q_{2} m \ell)} \right]
\end{equation}

The interaction energy for separated dislocation distributions is 
obtained from the cross terms (i.e., no self-energies) in Eq. 
(\ref{eq:energy1}).  Following the procedure outlined above, one first
calculates the interaction energy per unit length of dislocation line 
\begin{equation}
\label{eq:interact}
e_{int} = \frac{N \mu b^{2}}{\pi \; (1-\nu) \ell} 
\sum_{m=-\infty}^{\infty} \; 
 \int dq_{1} \; \frac{(2\pi m/\ell)^{2}}
{\left(q_{1}^{2}+(2 \pi m/\ell)^{2}\right)^{2}}\; \cos(q_{1} L) \; 
\exp{(-4\pi a|m|/\ell)}
\end{equation}
Upon evaluating this integral one finds that
\begin{equation}
\label{eq:e1int}
e_{int} = \frac{ N \mu b^{2}}{2 \; (1-\nu)} 
\sum_{m=1}^{\infty} \;\left[\left(\frac{L}{\ell} \right)+ \left(\frac{1}{2 
\pi |m|}\right)  \right] \exp{(-\alpha \; |m|)} , 
\end{equation}
where $\alpha = (4 \pi a/\ell) + (2 \pi L/\ell)$.
The resulting summations can also be performed to yield the 
grain-boundary interaction energy per unit area
\begin{equation}
\label{eq:gbint}
\gamma_{int} = \frac{e_{int}}{N \ell} = \frac{\mu b^{2}}{ 2 \pi (1-\nu) \ell} \; 
\left[\left(\frac{2 \pi L}{\ell} \right) \; \left(\frac{1}{\exp{(\alpha})-1} 
\right) - \ln{\left[1-\exp{(-\alpha)}\right]} \right]
\end{equation}
%
%

It is of interest to examine $\gamma_{int}$ for two limiting cases.  
First, for large boundary separations such that $L/\ell>>1$  and 
$a/\ell<<1$, $e_{int} \propto (L/\ell) \; \exp{(-2 \pi L/\ell)}$.  This exponential 
decay in the interaction energy follows from the rapid (exponential) 
decay of the stress fields associated with the individual arrays. By 
contrast, for $L/\ell \sim 1$, the logarithmic dependence of $e_{int}$ on $L/\ell$ 
results from individual dislocations in either array \lq\lq seeing" 
each other. Finally, we note that an alternative route to Eq. (\ref{eq:gbint}) 
follows from a calculation of the Peach-Koehler force acting on a 
dislocation owing to a distant array and a subsequent spatial integration to 
obtain the energy.\cite{tkl04} 

%
%
\subsection{Crack Array}
Another application of the formalism presented above is the
interaction energy between cracks. We first consider a single crack of 
length $2c$ oriented along the $r_{1}$ axis, and model it as 
a continuous dislocation distribution, with corresponding dislocation 
function $B(r_{1})$. \cite{weertman}  The Burgers vectors of the dislocations that 
comprise the crack model embody the local crack opening that results 
from a given loading.  Hence, by choosing an appropriate dislocation 
distribution, one can represent loading in various modes.  Having 
represented the crack with a dislocation distribution, the stress
field associated with the crack is given in terms of the stress field
of a single edge dislocation $\sigma_{ij}^{\perp}$ by the convolution integral
\begin{equation}
\label{eq:stresscrack}
\sigma_{ij}(r_{1},r_{2}) = \int_{-\infty}^{+\infty} dr_{1}^{\prime} \; 
B(r_{1}^{\prime}) \; \sigma_{ij}^{\perp}(r_{1}-r_{1}^{\prime},r_{2}) 
\end{equation}
As a specific example, consider a Mode II crack.  The shear loading 
associated with this mode can be represented by a distribution of edge 
dislocations with Burgers vectors oriented along $r_{1}$.  Thus, one can write 
$B(r_{1}) = \int dr_{2} \; \rho_{31}(r_{1},r_{2})$ for an appropriate 
dislocation density $\rho_{31}$ that can be determined from the 
requirement that the crack faces must be traction-free.\cite{weertman}

It is of interest here to consider a crack interacting with distant 
objects.  In particular, given an observation point $\vec{r}$ such that 
$|\vec{r}|/2c>>1$, it is permissible to regard the crack fields as 
produced by the lowest-order multipole moments of $B(r_{1})$.  In 
practice, the \lq\lq monopole" and \lq\lq dipole" moments are often 
satisfactory in this context and one obtains
\begin{equation}
\label{eq:moment1}
\int_{-\infty}^{+\infty} dr_{1} \; B(r_{1}) = b_{tot}, 
\end{equation}
and,
\begin{equation}
\label{eq:moment2}
D = \int_{-\infty}^{+\infty} dr_{1} \; r_{1} \; B(r_{1}) =  
\frac{-2 \left(1-\nu\right)}{\mu}\;\int_{-c}^{+c} dr_{1} \;
\sqrt{c^{2} - r_{1}^{2}} \;\; \sigma_{12}(r_{1},r_{2}=0),
\end{equation}  
where $b_{tot}$ is the total Burgers vector associated with the crack 
dislocations and $\sigma_{12}(r_{1},r_{2}=0)$ is the shear stress loading 
the crack.\cite{weertman,othercrack}

\begin{figure}
\includegraphics[scale=0.4]{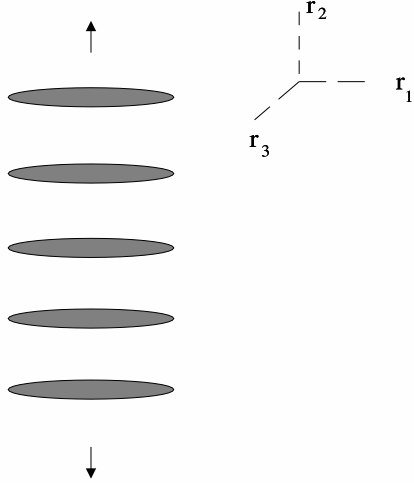}
\caption{An array of cracks, each separated from its nearest 
neighbor by a distance $\ell$ along the $r_{2}$ axis.  It is assumed 
that the system is subjected to a constant shear stress $\tau$.  Also 
shown is a schematic of the dislocation distribution corresponding to 
each crack.}
\label{fig:fig2}
\end{figure}

We now derive the interaction energy associated 
with various crack arrays.  Consider, for example, a linear array of Mode 
II cracks, each crack closed at both ends (i.e., $b_{tot}=0$) and 
modeled as a dislocation dipole in the $r_{2}$ direction that is separated 
from its nearest neighbor by 
$\ell$.  As is evident from Eq. (\ref{eq:moment2}), and by contrast with the 
grain-boundary models discussed above, the strength of elemental 
defects (i.e., their dipole moment) is not fixed, as it depends on the
local stress field.  The geometry for this loading, as 
well as the corresponding dislocation orientations, is shown in Fig. \ref{fig:fig2}.
In the short crack limit, such that each 
crack may be regarded as a point dipole with moment $D_{\ell}$ (i.e.,
the limit $c \rightarrow 0$ and $b \rightarrow \infty$ with $2cb
\rightarrow  D_{\ell}$), the corresponding dislocation density tensor is
\begin{equation}
\label{denscrack}
\rho_{31}(\vec{r}) = D_{\ell} \delta_{i3} \; \delta_{j1} \; 
\delta^{\prime}(r_{1}) \; \sum_{n} \delta(r_{2}-n\ell), 
\end{equation}
where the prime denotes differentiation with respect to the argument, 
and the subscript $\ell$ indicates that the dipole moment depends on 
crack separation. 
As before, it is convenient to work in reciprocal space where one 
finds that 
\begin{equation}
\label{eq:denscrackft}
\tilde{\rho}_{31}(\vec{q})  =  2 \pi i D_{\ell} \; q_{1} \; \delta(q_{3}) \; 
\sum_{n} \exp{( i q_{2} n \ell )}
\; \delta_{i3} \; \delta_{j1 }
\end{equation}

By analogy with the development given above, the energy of this system
is given by
\begin{equation}
\label{eq:cracken}
e_{c} = \frac{N \mu D_{\ell}^{2}}{(2\pi)\; (1-\nu) \ell} 
\sum_{m=-\infty}^{\infty} 
 \int dq_{1} \; q_{1}^{2} \; \frac{(2\pi m/\ell)^{2}}
{\left(q_{1}^{2}+(2 \pi m/\ell)^{2}\right)^{2}}
\exp{(-4\pi a|m|/\ell)}. 
\end{equation}
Upon performing the required integral and summation, one finds that
\begin{equation}
\label{eq:ecfinal}
e_{c} = \frac{ 2\pi \; N \mu D_{\ell}^{2}}{(1-\nu)\ell^{2}} \; 
\frac{\exp{\left(4 \pi a/\ell \right)}}{\left[\exp{\left(4 \pi a/\ell 
\right)}-1 \right]^{2}}.
\end{equation}
It is of interest to examine $e_{c}$ in the limit of large crack 
separations $\ell$ where the dipole approximation works best.  One
finds that $e_{c} \rightarrow \left[N \mu 
D_{\ell}^{2}/8 \pi (1-\nu) \right]\; \left[1/a^{2} - 4 \pi^{2}/3 \ell^{2} + 
O(1/\ell^{4}) \right]$, the first term arising from the self-energy 
of a dislocation dipole, and the second arising from dipole-dipole 
interactions.

An approximation to $D_{\ell}$ for a particular array spacing $\ell$
may be obtained by requiring that the shear 
stress traction produced by neighboring cracks on the face of a given crack 
be compensated by a uniform stress that is the same for every crack in 
the array.  The strategy for such calculations is given elsewhere
\cite{othercrack}, and the details for this defect geometry are given 
in Appendix $\ref{app:one}$.

\subsection{Crack-Grain Boundary Interaction}

\begin{figure}
\includegraphics[scale=0.4]{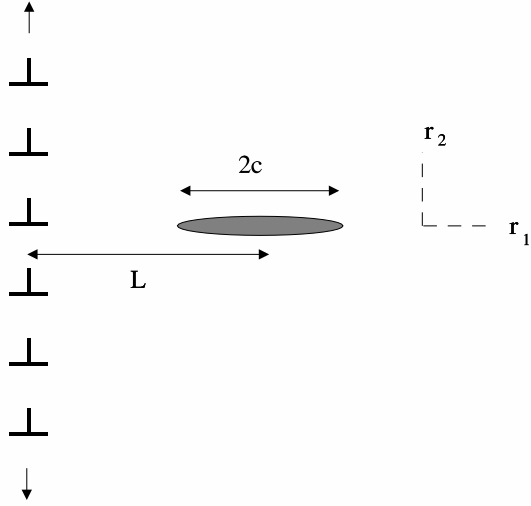}
\caption{An isolated crack interacting with a low-angle tilt 
boundary, the latter modeled as an array of edge dislocations.  For 
simplicity, the crack is located at a distance $L$ from the boundary 
and at $r_{2}=0$.  For this geometry the crack is subjected to shear 
stresses owing to the grain boundary.}
\label{fig:fig3}
\end{figure}

As a final illustration of extended defect energy calculations, consider 
the interaction of a low-angle grain boundary separated from an 
isolated crack by a distance $L$, as shown schematically in Fig. $\ref{fig:fig3}$.  
For this geometry the normal stress on the crack faces owing to the 
boundary is $\sigma_{22}(r_{1},r_{2}=0) = 0$, 
while the shear stresses $\sigma_{12}(r_{1},r_{2}=0) \neq 0$, and so the crack 
is loaded in Mode II.  From the 
grain-boundary (Eq. (\ref{eq:denstiltft})) and 
crack (Eq. (\ref{eq:denscrackft})) dislocation densities one 
can then construct the interaction energy per unit dislocation length, namely
\begin{equation}
\label{eq:interen}
e_{cgb} = \sum_{n}\frac{\mu b D_{L}}{(2\pi)^{2}\; (1-\nu)} \int \int d^{2}q 
\; \frac{q_{2}^{2}} {\left(q_{1}^{2}+q_{2}^{2}\right)^{2}}\; i \; 
q_{1} \; \exp{\left(i q_{1} L \right)} \cos{\left(q_{2} n \ell\right)} 
\exp{(-a|q_{2}|)}
\end{equation}

Evaluating the integrals and the summation, we find that  
\begin{equation}
\label{eq:interenfin}
e_{cgb} = \frac{\mu b D_{L}}{(2\pi)^{2}\; (1-\nu)} \; 
\frac{L \pi^{3}}{\ell^{2}} \; {\rm csch}^{2}\left[\pi\left(\frac{a+L}{L} 
\right)\right].
\end{equation}
In the limit of $\ell/L <<1$, $e_{cgb} \simeq \left(L/\ell^{2}\right)
    \; \exp{ \left[-2 \pi \left(a+L\right)/L \right]}$, owing to the
exponential decay of the stress fields associated with the grain
boundary while, for $\ell/L \simeq 1$ (and $a/L << 1$) 
$e_{int} \simeq 1/L$, consistent with a monopole-dipole interaction.  We 
note that this result can be obtained somewhat more readily by 
returning to the calculation of the grain boundary interaction energy 
and considering a single dislocation within a grain boundary array.  
Since a crack is modeled here as a dislocation dipole one replaces, 
in effect, an edge dislocation density with its spatial derivative and 
therefore the functional form of the crack/boundary interaction energy 
is given by $L \; \left(\partial e_{int}/\partial L\right)$ (see Eq. 
(\ref{eq:gbint})). 

The crack dipole moment $D_{L}$ can again be determined from the 
shear loading conditions via Eq. (\ref{eq:moment2}).  For the 
crack geometry shown in Fig. $\ref{fig:fig3}$, this leads to
\begin{equation}
\label{eq:stresstiltsxy}
D_{L} = \frac{ -2 b \pi}{ \ell^{2}} \;\int_{-c}^{+c} dr_{1} \; \
\sqrt{c^{2} - r_{1}^{2}} \; \left(r_{1}+L\right) \;
\left[\cosh{\left(\frac{2 \pi \left(r_{1}+L\right)}{\ell} \right)-1} 
\right]^{-1} ,
\end{equation}
%
%
where the shear stress is that for an array of edge dislocation. 
\cite{hirth}  This integral can be evaluated numerically by a
Gauss-Chebyshev integration technique based on type II Chebyshev 
polynomials. \cite{davis84}

\section{Conclusions}
A unified framework for the calculation of defect energies has been
presented here and, for the purposes of illustration, applied 
to several different systems containing grain boundaries and/or 
cracks. In each case, a set of elemental dislocations comprise the extended 
defect under consideration, and the corresponding dislocation density 
is either independent of other defect positions (e.g., grain boundary) 
or calculable from the stresses imposed by these other defects (e.g., 
crack).  As indicated above, this approach is especially 
useful in providing an intuitive understanding of interactions based
on  idealized defect models.

The foregoing development can, of course, be applied to other systems 
for which a dislocation-based model is appropriate.  For example, one 
can also obtain the energetics of a roughened tilt boundary 
\cite{rottmanprl86,rottman86} by considering 
perturbations of the dislocation density resulting from sinusoidal variations 
in the position of constituent dislocation lines.  If these 
variations in boundary morphology are temperature induced, this 
analysis can be used to compute the statistical weights of 
perturbed boundary configurations to the free energy of the system 
and, consequently, to deduce a thermodynamic roughening temperature.  
In addition, the description of more complex, asymmetrical boundary structures, 
consisting of two or three sets of edge dislocations, or twist 
boundaries consisting of perpendicular sets of screw dislocations is 
also possible by a generalization of the structure factor (see below 
Eq. (\ref{eq:enstrfac})) to include a form factor that reflects the positions of the 
basis dislocations within a unit cell that generates the boundary.  
These and other calculations will be the subject of a future publication.

Finally, we mention that, for the sake of completeness, Appendix
$\ref{app:stress}$ contains expressions for the stress
tensor in the same Fourier representation used here for the energy, along
with some simple examples.  In a future publication, we
will apply a similar formalism as used for the energy to derive equations for
the stress arising from  extended defects.

\section*{Acknowledgments}  The authors would like to acknowledge many
helpful discussion with Professor T. Delph and Mr. C. Lowe. This
research has been supported in part (JV) by the U.S. Department of
Energy under contract No. DE-FG05-95ER14566. The work of  R. LeSar was
performed under the auspices of the United States Department of Energy
(US DOE under contract W-7405-ENG-36) and was supported by the
Division of Materials Science of the Office of Basic Energy Sciences
of the Office of Science of the US DOE.

\appendix

\section{Crack arrays}
\label{app:one}
An approximation to the stress on a crack needed to balance those owing to others 
in the linear array shown in Fig. $\ref{fig:fig2}$ can be obtained via the formalism 
of Dyskin and M\"{u}hlhaus.\cite{othercrack}  We follow their 
approach below, except that the required stress fields are obtained 
by a multipole expansion rather than by using the Muskhelishvili complex 
potentials.\cite{musk}
First, assuming that the crack separation $\ell$ is large, one can 
regard each crack as a dislocation dipole.  For this geometry the 
relevant stresses associated with each crack, in the limit of small 
crack width $2c$, are given in terms of the derivatives of the 
%
%
derivative of the stresses (i.e., a point dipole approximation) for
individual dislocations by
\begin{equation}
\label{eq:sxy}
\sigma_{12}\left(r_{1},r_{2}\right) \approx \frac{\mu b c}{\pi 
\left(1-\nu\right)} \;
\frac{\partial}{\partial r_{1}}\left[\frac{r_{1} 
\left(r_{1}^{2}-r_{2}^{2}\right)}{\left(r_{1}^{2}+r_{2}^{2}\right)^{2}} 
\right]
= \frac{\mu D}{2 \pi 
\left(1-\nu\right)}
\;\left[\frac{3r_{1}^{2}-r_{2}^{2}}{\left(r_{1}^{2}+r_{2}^{2} \right)^{2}} - 
\frac{4 r_{1}^{2} 
\left(r_{1}^{2}-r_{2}^{2}\right)}{\left(r_{1}^{2}+r_{2}^{2}\right)^{3}}\right],
\end{equation}
and
\begin{equation}
\label{eq:syy}
\sigma_{22}\left(r_{1},r_{2}\right) \approx \frac{\mu b c}{\pi 
\left(1-\nu\right)} \; \frac{\partial}{\partial r_{1}} \left[\frac{r_{2} 
\left(r_{1}^{2}-r_{2}^{2}\right)}{\left(r_{1}^{2}+r_{2}^{2}\right)^{2}} 
\right] = \frac{\mu D \;r_{1} r_{2}}{\pi \left(1-\nu\right)} \;  
\left[\frac{1}{\left(r_{1}^{2}+r_{2}^{2}\right)^{2}} - 
\frac{2 
\left(r_{1}^{2}-r_{2}^{2}\right)}{\left(r_{1}^{2}+r_{2}^{2}\right)^{3}} \right],
\end{equation}
where $D$ is the dipole moment.  Given the symmetry of the crack
array, it is clear that the 
stresses $\sigma_{22}$ will cancel upon summation over all cracks.

Next, we invoke a dipole asymptotics approximation \cite{othercrack} 
in which the $i$-th crack is subjected to a loading shear stress $\tau$ 
and an additional uniform load $\epsilon_{i}$, the latter equal to the stresses 
generated by the other cracks at the center of the $i$-th crack.  The 
corresponding dipole moment is $D_{i} = -\pi c^{2} \left( 1-\nu \right) 
\left(\tau+\epsilon_{i} \right)/\mu$.  The corrective stress associated 
with the $j$-th crack is then given by
\begin{equation}
\label{eq:systemeq}
\epsilon_{j} = \frac{c^{2}}{2 \ell^{2}} \; \sum_{i \neq j} 
\frac{1}{\left( i-j\right)^{2}} \; \left(\tau+\epsilon_{i} \right).
\end{equation}
Finally, taking each crack to be identical so that $\epsilon_{i} = 
\epsilon$ and using the Riemann zeta function relation
$\sum_{n=1}^{\infty} 1/n^{2} = \zeta(2)=\pi^{2}/6$,\cite{arfken} one finally obtains
\begin{equation}
\label{eq:epsfin}
\epsilon = \tau \; \frac{\left(c/\ell\right)^{2} 
\zeta(2)}{1-\left(c/\ell\right)^{2} \zeta(2)}.
\end{equation} 
The dipole moment can now be expressed in terms of $\tau$ and $\epsilon$.

\section{Stress Tensor}\label{app:stress}

The stress tensor can also be given in terms of the dislocation
density tensor. As discussed by Kosevich \cite{kosevich}, the stress
field at point $\vec{r}$ 
\begin{equation}
\label{eq:stress}
\sigma_{ik}(\vec{r}) = 2 \mu \left[ \nabla^{2} \chi_{ik}^{\prime} +
  \frac{1}{1-\nu}  \left( \frac{\partial^{2}
  \chi_{ll}^{\prime}}{\partial x_{i} \partial x_{k}} - \delta_{ik}
  \nabla^{2} \chi_{ll}^{\prime} \right) \right] ,
\end{equation}
with
\begin{equation}
\label{eq:bihar_green}
\chi_{ik}^{\prime} = - \frac{1}{8 \pi} \int |\vec{r}-\vec{r}^{\prime}|
\eta_{ik}(\vec{r}^{\prime}) d\vec{r}^{\prime},
\end{equation}
and
\begin{equation}
\label{eq:eta}
\eta_{ik} = \frac{1}{2} \left[ \epsilon_{ipl} \frac{\partial
    \rho_{kl}}{\partial x_{p}} +\epsilon_{kpl} \frac{\partial
    \rho_{il}}{\partial x_{p}} \right].
\end{equation}

In Fourier space one can write Eq. (\ref{eq:bihar_green}) as
\begin{equation}
\label{eq:biharfour}
\tilde{\chi}_{ik}^{\prime}(\vec{q}) = \tilde{G}(\vec{q}) \;
\tilde{\eta}_{ik}(\vec{q}),
\end{equation}
with $\tilde{G}(\vec{q})= 1 /q^{4}$, the Green function of the
biharmonic operator \cite{footnote}. Furthermore,
\begin{equation}
\label{eq:etafour}
\tilde{\eta}_{ik}(\vec{q}) = \frac{i}{2} \left[ \epsilon_{ipl} q_{p}
  \rho_{il}(\vec{q}) + \epsilon_{kpl} q_{p} \rho_{il}(\vec{q}) \right]
  = - \frac{i q}{2} \left[ C_{il} \rho_{kl}(\vec{q}) + C_{kl}
  \rho_{il}(\vec{q}) \right].
\end{equation}
Given these transforms, the Fourier transform of the stress tensor can
be written, after some algebra, as 
\begin{equation}
\label{eq:stressfour}
\tilde{\sigma}_{ik}(\vec{q}) = i \mu q^{3} \tilde{G}(\vec{q}) \left(
C_{il} \rho_{kl} + C_{kl} \rho_{il} - \frac{2}{1-\nu} Q_{ik}
C_{al}\rho_{al} \right).
\end{equation}

To illustrate the use of Eq. (\ref{eq:stressfour}), consider the case
of a single straight screw dislocation, aligned along $r_{3}$ and
having Burgers vector $\vec{b}$. The corresponding dislocation density
is $\tilde{\rho}_{kl} = 2 \pi \; b \; \delta_{k3} \delta_{l3} \;
\delta(q_{3})$.

Upon substituting this density into Eq. (\ref{eq:stressfour}) one
finds that 
\begin{equation}
\label{eq:sigmascrew}
\tilde{\sigma}_{ik}(\vec{q}) = 2 \pi i \mu q^{3} \;
\tilde{G}(\vec{q})\;  b \; \delta(q_{3})  \left(C_{i3} \delta_{k3}
+C_{k3} \delta_{i3} \right).
\end{equation}
Therefore, one immediately sees that $\tilde{\sigma}_{11} =
\tilde{\sigma}_{22} = \tilde{\sigma}_{33} = 0$, and that
$\tilde{\sigma}_{12}=0$. The non-vanishing stress component
\begin{equation}
\label{eq:s13four}
\tilde{\sigma}_{13} = - i 2 \pi \mu b \;
\left(\frac{q_{2}}{q^{2}}\right) \; \delta(q_{3}).
\end{equation}
and so, in real space, 
\begin{equation}
\label{eq:s13} \sigma_{13} = - \frac{\mu b y}{2 \pi r^{2}}.
\end{equation}
A similar result can be derived for $\sigma_{23}$.


\begin{thebibliography}{99}
\bibitem{footnote} The 
Green function of the biharmonic operator can be obtained by noting 
that, in infinite space,
$\nabla^{4} |\vec{r}| = \nabla^{2} \nabla^{2} |\vec{r}| = -
\delta(\vec{r})$.  This result follows from the Green function 
relation for the three-dimensional Laplace
operator $\nabla^{2} | \vec{r} | = - \frac{1}{|\vec{r}|}$.
\end{thebibliography}
\end{document}